\documentclass[10pt,twocolumn,superscriptaddress,showpacs,pre]{revtex4-1}  
\usepackage{amsmath}
\usepackage{verbatim}
\usepackage{graphicx}
\usepackage{subfigure}

\begin{document}

\title{\large{Synchronisation of stochastic oscillators in biochemical systems}}

\author{Joseph~D. Challenger}
\email{joseph.challenger@postgrad.manchester.ac.uk}
\affiliation{Dipartimento di Energetica, Universit\`{a} degli Studi di Firenze, via Santa Marta 3, 50139 Florence, Italy}
\affiliation{Theoretical Physics Division, School of Physics and Astronomy, 
The University of Manchester, Manchester M13 9PL, UK}
\author{Alan~J. McKane}
\email{alan.mckane@manchester.ac.uk}
\affiliation{Theoretical Physics Division, School of Physics and Astronomy, 
The University of Manchester, Manchester M13 9PL, UK}

\begin{abstract}
A formalism is developed which describes the extent to which stochastic 
oscillations in biochemical models are synchronised. It is based on the 
calculation of the complex coherence function within the linear noise 
approximation. The method is illustrated on a simple example and then applied 
to study the synchronisation of chemical concentrations in social amoeba. The 
degree to which variation of rate constants in different cells and the volume
of the cells affects synchronisation of the oscillations is explored, and the phase lag calculated. In all
cases the analytical results are shown to be in good agreement with those 
obtained through numerical simulations.
\end{abstract}
\maketitle

\section{Introduction}
\label{sec:intro}

Oscillatory behaviour is observed in a great variety of biochemical systems, 
over a wide range of time periods {\cite{hess_71,goldbeter_96}}. These 
oscillations have been modelled extensively, using both deterministic and 
stochastic frameworks {\cite{lefever_67,gonze_02,mckane_07,wang_08}}. Most of 
this modelling is carried out at the single cell level. However, many processes
display coherent behaviour over a population of cells. This requires that the 
oscillations in the individual cells must influence each other. One reason for 
this is that random fluctuations, intrinsic to these systems, can introduce 
random phase lags. This means that a population of isolated cells demonstrating
oscillatory dynamics will not remain synchronised over time, even if they are 
synchronised initially. Therefore, for coherent, collective behaviour (such 
as intercellular signalling), some form of communication between the cells 
is necessary. 

Before proceeding further we need to discuss what is meant by the term 
`synchronisation'. This is necessary to do this since its precise definition 
varies. For the stochastic time series of the kind that will interest us here, 
the approach of Pikovsky, Rosemblum \& Kurths will be the most relevant. They 
define synchronisation as ``an adjustment of rhythms of oscillating objects 
due to their weak interaction''{\cite{pikovsky_01}}. Using reaction systems 
as examples, we will show the forms that the interaction can take, and what 
`weak' means in this context. 

One place in which the role and mechanisms of synchronisation have been 
systematically analysed is the study of the dynamics of glycolytic metabolism. 
Richard \textit{et.~al.} have performed experiments in yeast cells, and 
observed oscillations both in individual cells and populations of cells 
{\cite{richard_96,richard_96b}}. The authors conclude that the coupling of 
the cells \textit{via} an extracellular metabolite (proposed to be 
acetaldehyde) causes the oscillations of the individual cells to synchronise. 
In {\cite{richard_96b}}, two cell populations, originally oscillating 
$180^{\circ}$ out of phase, became synchronised upon coupling, and a time to 
synchronise was measured.

Several mathematical models have attempted to capture and explain this 
behaviour. One of these is due to Wolf \& Heinrich, devised in {\cite{wolf_97}}
and refined in {\cite{wolf_00}}. They begin by describing a deterministic, 
one-cell model, where the cell is connected to an extra-cellular compartment, 
which exhibits either a stable fixed point or a stable limit cycle, depending 
on the values of the reaction parameters. Wolf \& Heinrich then extended the 
model to $n$-cells, where the cells are coupled by exchanging molecules with 
a shared extra-cellular compartment. In {\cite{wolf_00}}, the authors show 
that in a two-cell model, synchronisation with either zero-phase or non-zero 
phase lag is observed, depending on the reaction parameters chosen. Following 
the experiments by Richard \textit{et.~al.}, the authors also mixed two populations of 
cells which, before mixing, were internally synchronised but out of phase with 
the other population. Wolf \& Heinrich found that, once mixed, the two 
populations did indeed synchronise, although this took considerably longer 
to achieve than it did in the experiments. This may suggest that the form of 
the coupling, or the chemical species chosen to be responsible for the 
coupling, may not be the correct one. In many models it has been found that 
the form of the coupling chosen in the model can significantly alter the 
dynamics observed, as found in {\cite{wang_05,gonze_08}}.

Another system in which synchronisation is often studied involves oscillations 
in the concentration of adenosine 3', 5'-cyclic monophosphate ($cAMP$) in the 
amoeba \textit{Dictyostelium discoideum} {\cite{goldbeter_96,kamino_11}}.  
These social amoeba feed on bacteria in the forest soil. During the onset of 
starvation, the cells alter their behaviour to aggregate, and become able to 
relay $cAMP$ signals, in the form of oscillations. In this way, populations of 
the \textit{Dictyostelium} cells form large aggregates. This process culminates
with the formation of a fruiting body which can disperse
spores, leading back to the start of the life cycle  {\cite{goldbeter_96}}. One reason why this is 
much studied is that many of the components involved in the $cAMP$ signalling 
have corresponding components in mammalian pathways, hence the desire to 
understand this system in more detail \cite{williams_06}. 

In this paper we will present an approach to the investigation of 
synchronisation in models of biochemical processes which is analytic and where
the models are stochastic in nature. The vast majority of studies of such 
systems found in the literature are described by systems of ordinary 
differential equations (ODEs), which show limit cycle behaviour. However it is
now well established that stochastic effects will be significant even in 
systems containing a large number of molecules 
\cite{elf_03b,kummer_05,mckane_07}. Stochastic oscillations will not only 
occur at a single frequency, as for limit cycles, and this will have 
consequences when investigating synchronisation of these oscillations. What 
studies there have been of synchronisation of stochastic oscillations have 
been purely numerical \cite{kim_07}, and have not considered the possibility 
of phase lags between oscillations in different cells. However, the key aspects
of synchronisation in stochastic biochemical models become particularly clear 
when an systematic analytical treatment is given. 

The outline of the paper is as follows. In Section \ref{sec:method} we will 
introduce the formalism we shall be using and give a simple example to 
illustrate its use. A model of $cAMP$ oscillations in \textit{Dictyostelium 
discoideum} will be analysed in Section \ref{sec:dict}, to show the application
of our methods to a more realistic, multi-cell model. In particular, we show 
under which conditions a phase lag between oscillations in neighbouring cells 
is introduced. We end with a summary and suggestions for further work.

\section{Methodology}
\label{sec:method}
In this section we will describe the formulation of the process in terms of a
stochastic model, and then go on to describe the methods we use to see if
synchronisation occurs. Fortunately, the quantities we need to calculate to
investigate the question of synchronisation are the same as those required 
to study stochastic oscillations, and so in many cases much of the calculation will 
already have been carried out.

\subsection{Formalism}
\label{sec:formalism}
The reaction system is described by a chemical master equation, which is 
written in terms of the integer populations of the chemical species present 
in the model, which we write as $\boldsymbol{n}=(n_1,n_2,...,n_K)$, if there 
are $K$ species present in the model. This describes the time evolution of 
the probability for the system to be in state $\boldsymbol{n}$ at time $t$, 
which is written as $P_{\boldsymbol{n}}(t)$. Transitions from one state to 
another are caused by the chemical reactions. We write 
$T(\boldsymbol{n'}|\boldsymbol{n})$ as the transition rate from state 
$\boldsymbol{n}$ to state $\boldsymbol{n}' = \boldsymbol{n}-\boldsymbol{\nu}$, 
where $\boldsymbol{\nu}_{\mu} = (\nu_{1 \mu},\ldots,\nu_{L \mu})$ is the
stoichiometric vector corresponding to reaction $\mu$. The master equation then
has the general form \cite{kampen_07,pahle_12}
\begin{equation}
\begin{split}
\frac{\mathrm{d}P_{\boldsymbol{n}}(t)}{\mathrm{d}t} = \sum_{\mu}
%\left[ 
T_{\mu}(\boldsymbol{n}|\boldsymbol{n}-\boldsymbol{\nu}_{\mu})
P_{\boldsymbol{n}-\boldsymbol{\nu}_{\mu}}(t) \\
-T_{\mu}(\boldsymbol{n}+\boldsymbol{\nu}_{\mu}|\boldsymbol{n})
P_{\boldsymbol{n}}(t) 
%\right].
\end{split}
\label{master}
\end{equation}

To make progress analytically, an approximation scheme must be employed. The
crudest scheme is to ignore fluctuations completely, and simply write down
an equation for the average 
\begin{equation}
\langle n_i(t) \rangle \equiv \sum_{\boldsymbol{n}} n_i P_{\boldsymbol{n}}(t), \ \ 
i=1,\ldots,K.
\label{average}
\end{equation}
In the limit where both the number of molecules in the system and its volume, 
$V$, becomes large, we describe the system in terms of the concentrations of
the chemical species, $x_{i}=\lim_{V \to \infty}\langle n_i \rangle/V$. From 
Eqs.~(\ref{master}) and (\ref{average}) one finds the macroscopic, 
deterministic, equation for $x_i(t)$:
\begin{equation}
\frac{\mathrm{d}x_i}{\mathrm{d}t} = A_i(\boldsymbol{x}), \ \ \ \ \ \ \
A_i(\boldsymbol{x}) \equiv \sum_{\mu} \nu_{i \mu}f_{\mu}(\boldsymbol{x}), 
\ \ \ i=1,\ldots,K,
\label{macroscopic_eqn}
\end{equation}
where $f_{\mu}(\boldsymbol{x})$ is defined by \cite{pahle_12}
\begin{equation}
f_{\mu}(\boldsymbol{x}) = \lim_{V \to \infty} 
T_{\mu}\big(\langle\boldsymbol{n}\rangle+
\boldsymbol{\nu}_{\mu}|\langle\boldsymbol{n}\rangle\big)/V.
\label{f_defn}
\end{equation}
It is straightforward to calculate the function $A_i(\boldsymbol{x})$ from 
the chemical reactions and the reaction rates, and so we can determine whether 
at large times the system tends to a fixed point, a limit cycle, or some other,
more complex, type of behaviour.

However, as mentioned already, oscillations seen in biochemical reactions may
be stochastic in nature, and so not be described by the deterministic equation
(\ref{macroscopic_eqn}). To study these we cannot completely ignore the dynamics
of the fluctuations contained in Eq.~(\ref{master}). The simplest way of doing
this is to simply include stochastic effects (or `noise') as linear deviations
from the deterministic result. This is the linear noise approximation (LNA)
and is especially simple to implement if the system has reached a stationary
state which corresponds to a fixed point of the deterministic equation 
(\ref{macroscopic_eqn}). The LNA then consists of working with the variables 
$x_i(t)$, writing them as $x_i = x_i^{*} + V^{-1/2}\xi_i$. The LNA assumes that $V$ is large, so we keep only linear terms in $\xi_i$ when they are substituted into the master equation 
(\ref{master}). Here the asterisk signifies a fixed point of 
Eq.~(\ref{macroscopic_eqn}) and the $V^{-1/2}$ reflects the Gaussian nature of
the approximation \cite{kampen_07}. The equation satisfied by the linearised 
fluctuations $\xi$ is found to be (for details, we refer the reader to the 
literature \cite{elf_03b,mckane_07,pahle_12})
\begin{equation}
 \frac{\textrm{d}\xi_i}{\textrm{d}t}=\sum_{j=1}^{K} M_{ij}\xi_j + \eta_i(t), 
\ \ \ \ i=1,\ldots,K,
\label{langevin}
\end{equation}
where $\eta_i(t)$ is a Gaussian noise term with zero mean and with correlator 
$\langle \eta_i(t) \eta_j(t') \rangle=B_{ij}\delta (t-t')$. The matrix $M$ is
the Jacobian at the fixed point and the matrix $B$ can be calculated in the 
same way that the function $A$ was:
\begin{equation}
\begin{split}
M_{i j} &= \sum_{\mu} \nu_{i \mu}\left. \frac{\partial f_{\mu}}
{\partial x_j} \right|_{\boldsymbol{x}=\boldsymbol{x}^{*}}, \\ 
B_{i j} &=\sum_{\mu} \nu_{i \mu} \nu_{j \mu} f_{\mu}(\boldsymbol{x}^{*}), \ \ 
i,j = 1,\ldots,K.
\end{split}
\label{MandB}
\end{equation}

Since Eq.~(\ref{langevin}) describes fluctuations about the stationary state and is linear,
we can analyse the equation using Fourier transforms. Denoting the temporal
Fourier transform of $\xi_i(t)$ as $\tilde{\xi}_i(\omega)$, we may write 
Eq.~(\ref{langevin}) as 
$\tilde{\xi}_i(\omega)=\sum_{j}\Phi^{-1}_{ij}(\omega)\tilde{\eta}_j(\omega)$ 
where $\Phi_{ij}\equiv -M_{ij}-i\omega \delta_{ij}$. This then allows us to 
calculate the power spectral density matrix (PSDM) 
\begin{equation}
P_{ij}(\omega) \equiv \langle \tilde{\xi}_i(\omega)
\tilde{\xi}^*_j(\omega) \rangle=\sum_{l=1}^K\sum_{m=1}^K 
\Phi_{il}^{-1}(\omega)B_{lm}(\Phi^{\dagger})^{-1}_{mj}(\omega).
\label{psdm}
\end{equation}
In earlier work on analysing the nature of the stochastic oscillations about a
fixed point, only the diagonal entries of this matrix (the `power 
spectra' $P_{ii}$) were of interest \cite{mckane_05,mckane_07}. Here we shall 
also be interested in the off-diagonal entries, but as mentioned in the 
Introduction, if Eq.~(\ref{psdm}) has been calculated so as to obtain the power
spectrum, the off-diagonal elements can be immediately read-off.

An application quite similar to the one we are discussing here is that 
described by Rozhnova \textit{et. al.} who used the off-diagonal elements to 
show how stochastic oscillations can become synchronised \cite{rozhnova_12}. 
They constructed an epidemiological model of a network of cities, where a 
disease can be transported between cities due to movement of infected 
commuters. We will use a similar approach but instead of cities and people, 
we will work with cells and molecules. Unlike the diagonal elements --- the 
power spectra --- the off-diagonal elements of the PSDM will in general be 
complex. Often, these elements are normalised, by using the relevant power 
spectra. This quantity is then known as the complex coherence function (CCF)
\cite{priestley_81,marple_87,rozhnova_12},
\begin{equation}
 C_{ij}(\omega)=\frac{P_{ij}(\omega)}{\sqrt{P_{ii}(\omega)P_{jj}(\omega)}}.
\label{ccf}
\end{equation}
 
As this is a complex function, it can be expressed in terms of a magnitude and 
a phase,
\begin{equation}
|C_{ij}(\omega)|=\frac{|P_{ij}(\omega)|}{\sqrt{P_{ii}(\omega)P_{jj}(\omega)}},
\label{mag_c}
\end{equation}
\begin{equation}
\phi_{ij}(\omega)=\textrm{arctan}[\frac{\textrm{Im}(C_{ij}(\omega))}
{\textrm{Re}(C_{ij}(\omega))}]=\textrm{arctan}[\frac{\textrm{Im}
(P_{ij}(\omega))}{\textrm{Re}(P_{ij}(\omega))}].
\label{phase}
\end{equation}
The magnitude of the CCF tells us the similarity between two signals, as a 
function of $\omega$. The phase of the CCF tells us the phase lag between two 
signals as a function of the frequency $\omega$ {\cite{marple_87}}. For 
example, if the two signals are in phase the CCF will be a real function. To
clarify the use of these concepts we apply them to a simple example.

\subsection{Simple example}
\label{sec:simple}
The biochemical reaction scheme we will use to illustrate the methodology is a
toy system, but many more complex reaction schemes contain molecular species which 
behave in this way. The system consists of two species $A$ and $B$, which may
both lose molecules through diffusion out of the cell and also gain them 
through diffusion into the cell. Species $A$ requires molecules of species 
$B$ to sustain its numbers, and they become depleted when too many $B$ molecules
have been used up. This is essentially a predator-prey dynamics, and we will
use this language to discuss it, since it makes the results more intuitive. The
sustained stochastic oscillations have been previously investigated using the
formalism described in Section \ref{sec:formalism} \cite{mckane_05}. The 
actions of these species is described by the following interactions and their 
corresponding transition rates: 

\begin{align}
B + E \stackrel{b}\longrightarrow B + B,\,\,\,\,   
\nonumber \\  T(n_1,n_2+1&|n_1,n_2)= bn_2\frac{(N-n_1-n_2)}{N} , \nonumber \\
A \stackrel{d}\longrightarrow E,\,\,\,\,    
&T(n_1-1,n_2|n_1,n_2)= dn_1, \nonumber \\
A + B \stackrel{p_1}\longrightarrow A + A,\,\,\,\,   
&T(n_1+1,n_2-1|n_1,n_2)= p_1\frac{n_1n_2}{N}, \nonumber \\
A + B \stackrel{p_2}\longrightarrow A + E,\,\,\,\,   
&T(n_1,n_2-1|n_1,n_2)=p_2\frac{n_1n_2}{N}.
\end{align}
The variable $E$ is introduced to give the system a maximum carrying capacity, 
which is the `size of the system' $N$. That is, the total number of individuals
at any time is fixed, $n_1+n_2+n_E=N$, where $n_1, n_2$ and $n_E$ are, 
respectively, the numbers of $A, B$ and $E$ molecules in the cell. Because of this we will use $N$ as the large parameter employed in the LNA in this example. We have also only 
allowed for species $A$ to diffuse out of the cell, and for species $B$ to 
diffuse in (subject to existing concentration). The relation $n_E=N -n_1 - n_2$
has been used to simplify the transition rates above. 

Applying the methodology described in Section \ref{sec:formalism}, the 
deterministic equations (\ref{macroscopic_eqn}) are
\begin{align}
\frac{\textrm{d}x_1}{\textrm{d}t}&=p_1x_1x_2-dx_1, \nonumber \\
\frac{\textrm{d}x_2}{\textrm{d}t}&=bx_2(1-x_1-x_2)-(p_1+p_2)x_1x_2.
\end{align}
To go beyond this, we use the LNA which is fully characterised by the two 
matrices $M$ and $B$, which can be calculated from Eq.~(\ref{MandB}), and 
which are given explicitly in \cite{mckane_05}.
\begin{comment}
which in this example are given by \cite{mckane_05}
\begin{equation}
M=\left( \begin{array}{cc} -d+p_1y & p_1x\\
-by-(p_1+p_2)y&-(p_1+p_2)x+b(1-x-2y)
\end{array} \right),
\end{equation}
\begin{equation}
B=\left( \begin{array}{cc} dx+p_1xy & -p_1xy \\
-p_1xy & by(1-x-y)+(p_1+p_2)xy
\end{array} \right).
\end{equation}
\end{comment}

\begin{figure}
\begin{center}
 \includegraphics{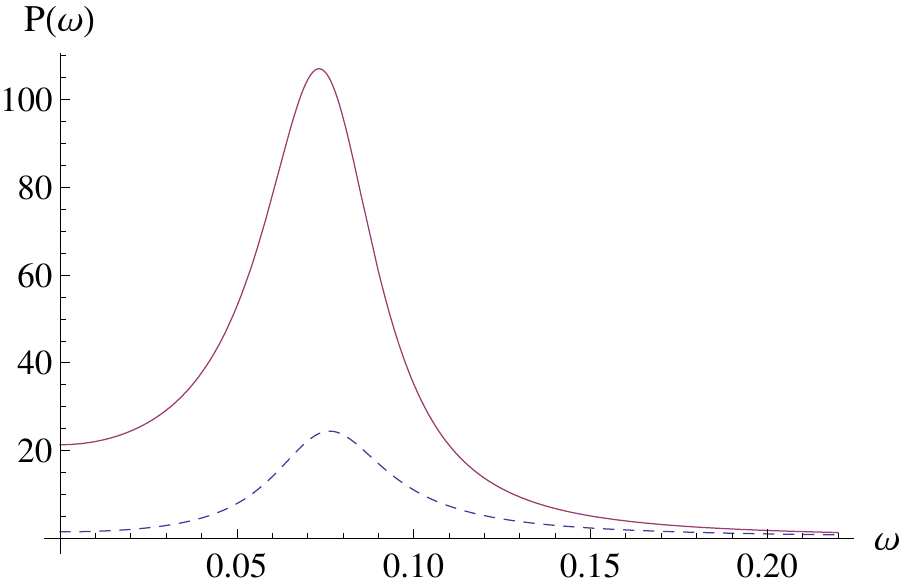}
\caption{Power spectra of the oscillations observed in the two-species 
example calculated using the LNA. The red (solid) curve is for the prey and 
the blue (dashed) curve for the predator. The parameter values used are: $b=0.1$, $d=0.1$, $p_1=0.25$, $p_2=0.05$, and $N=3200$.}
\label{fig:ppp}
\end{center}
\end{figure}

The macroscopic equations have a single nontrivial stable fixed point (or `steady state'), and for 
the reaction parameters chosen here the eigenvalues of the Jacobian, evaluated 
at the fixed point, have an imaginary component and sustained stochastic
oscillations can be observed around the fixed point \cite{mckane_05}. This is
illustrated in Figure~\ref{fig:ppp} which shows the power spectra of the 
oscillations, calculated using Eq.~(\ref{psdm}) for a particular choice of the reaction parameters. The large peak shows the 
existence of amplified stochastic oscillations for a reasonably narrow 
range of values about a characteristic frequency \cite{mckane_05}.

%%%%%%%%%%%%%%%%%%%%%%%%%%%%%%%%%%%%%%%%%%%%%%%%%%%%%%%%%%%%%%%%%%%%%%%%%%%
%%%%%%%%%%%%%%%%%%%%%%%%%%%%%%%%%%%%%%%%%%%%%%%%%%%%%%%%%%%%%%%%%%%%%%%%%%%

%%%%%%%%%%%%%%%%%%%%%%%%%%%%%%%%%%%%%%%%%%%%%%%%%%%%%%%%%%%%%%%%%%%%%%%%%%%
%%%%%%%%%%%%%%%%%%%%%%%%%%%%%%%%%%%%%%%%%%%%%%%%%%%%%%%%%%%%%%%%%%%%%%%%%%%

\begin{figure}
\begin{center}
 \includegraphics{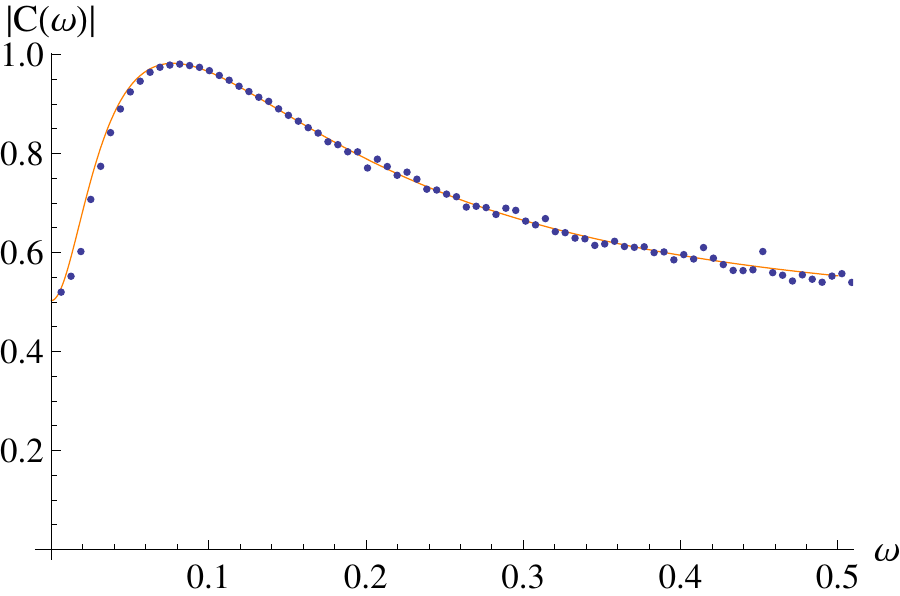}
\caption{Magnitude of the CCF $C_{12}(\omega)$ in the predator prey model. 
The orange curve is the theoretical prediction and the blue dots are the 
results obtained from numerical simulation.}
\label{fig:pp_ccf}
\end{center}
\end{figure}

\begin{figure}
\begin{center}
 \includegraphics{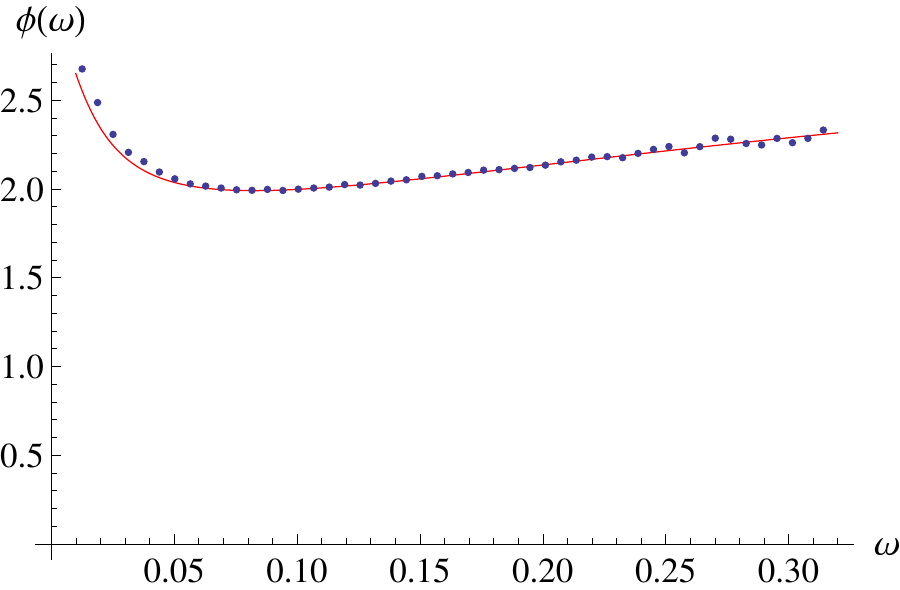}
\caption{The phase spectra in the predator prey model, showing the phase lag 
(in radians) between the oscillators as a function of $\omega$. The red line 
is the theoretical prediction and the blue dots are the results obtained from 
numerical simulation.}
\label{fig:pp_phase}
\end{center}
\end{figure}

We use the complex coherence function (CCF), introduced in the previous 
section, to study the relation between these two oscillators. In general, the 
CCF will be a complex-valued function. Figure~\ref{fig:pp_ccf} shows the 
magnitude of the CCF, showing a strong shared signal in the two oscillators. 
The analytical result obtained from Eq.~(\ref{mag_c}), is compared with 
results from numerical simulations, obtained using the Gillespie algorithm 
\cite{gillespie_76}. The phase lag, defined in Eq.~(\ref{phase}), is found to
have the simple form
\begin{equation}
\phi(\omega) = \tan^{-1}\left[ \frac{\alpha \omega}{\beta + 
\gamma \omega^2} \right],
\label{phase_pp}
\end{equation}
for this two species model. Here $\alpha, \beta$ and $\gamma$ are calculable 
combinations of the parameters $b, d, p_1$ and $p_2$ defining the system. The 
phase lag is shown in Figure~\ref{fig:pp_phase} along with simulation results. 
Figure~\ref{fig:ppp} indicates that the oscillations are most significant over 
the frequency range $0.05\leq \omega \leq 0.11$. 
The phase tends to $\pi$ as $\omega \to 0$ and 
$\omega \to \infty$. However within the range $0.05\leq \omega \leq 0.10$ the 
phase lag  remains fairly constant, at around 2 radians. 

We end this section by remarking that we do not use the word synchronisation
to describe the phenomena observed in this section. We will only use this term
for the interaction between self-sustained oscillators. We do not have two 
self-sustained oscillators here, as neither species would continue to oscillate
if the other were removed or held fixed. In the next section, we will look at 
a more realistic, multi-cellular model, where self-sustained oscillators in 
neighbouring cells become synchronised due to the coupling between the cells. This is what we mean by the `weak interaction' between oscillators.

%%%%%%%%%%%%%%%%%%%%%%%%%%%%%%%%%%%%%%%%%%%%%%%%%%%%%%%%%%%%%%%%%%%%%%%%%%%
%%%%%%%%%%%%%%%%%%%%%%%%%%%%%%%%%%%%%%%%%%%%%%%%%%%%%%%%%%%%%%%%%%%%%%%%%%%

%%%%%%%%%%%%%%%%%%%%%%%%%%%%%%%%%%%%%%%%%%%%%%%%%%%%%%%%%%%%%%%%%%%%%%%%%%%
%%%%%%%%%%%%%%%%%%%%%%%%%%%%%%%%%%%%%%%%%%%%%%%%%%%%%%%%%%%%%%%%%%%%%%%%%%%

%%%%%%%%%%%%%%%%%%%%%%%%%%%%%%%%%%%%%%%%%%%%%%%%%%%%%%%%%%%%%%%%%%%%%%%%%%%
%%%%%%%%%%%%%%%%%%%%%%%%%%%%%%%%%%%%%%%%%%%%%%%%%%%%%%%%%%%%%%%%%%%%%%%%%%%

%%%%%%%%%%%%%%%%%%%%%%%%%%%%%%%%%%%%%%%%%%%%%%%%%%%%%%%%%%%%%%%%%%%%%%%%%%%
%%%%%%%%%%%%%%%%%%%%%%%%%%%%%%%%%%%%%%%%%%%%%%%%%%%%%%%%%%%%%%%%%%%%%%%%%%%
 
\section{Collective Behaviour in \textit{Dictyostelium discoideum}}
\label{sec:dict}

The reaction system we will now discuss is a model of oscillations 
in the concentration of adenosine 3', 5'-cyclic monophosphate ($cAMP$), which 
are observed within the amoeba \textit{Dictyostelium discoideum}. The model is 
due to Kim~\textit{et.~al.} and in their work was used to study the synchrony 
of stochastic oscillations in the concentration of $cAMP$ across multiple cells
\textit{via} numerical simulation. Previously, the pathway that produces these 
oscillations had been modelled using a set of non-linear ODEs, which exhibit 
limit cycle behaviour {\cite{goldbeter_96}}. Starting from such a model, 
described in {\cite{laub_98}}, Kim~\textit{et.~al.} {\cite{kim_07}} show that 
considering stochastic oscillations, in addition to those in the limit cycle 
regime, increases the robustness of the oscillations. That is, oscillations, 
whether deterministic or stochastic in origin, are observed over a larger 
volume of parameter space than would be the case if only deterministic 
oscillations were considered. Kim~\textit{et.~al.} also demonstrate that, in 
the deterministic model, the oscillations cease to be observed if the reaction 
parameter values are slightly altered. The robustness of the oscillations in 
the model is desirable, as any parameter values present in a mathematical model
of a biochemical reaction system will have an associated uncertainty. In 
addition to this, because the physical system involves an aggregation of cells,
conditions (\textit{e.g.} temperature) may vary from cell to cell, so the 
reaction parameters may not be identical inside each cell. Because of this, 
Kim~\textit{et.~al.} made small perturbations to the reaction parameters in 
different cells to see if oscillations were still observed, and whether they 
still synchronised. 

In this section we will use the theoretical framework, outlined in Section 2, 
to obtain analytical results for the system in terms of the CCF. This provides 
information on the strength of coupling between the oscillations in 
neighbouring cells (through the magnitude of the CCF), as well as the phase 
lag between the oscillations, if any. The possibility of a phase lag, 
introduced due to the heterogeneity of the reaction parameters in different 
cells was not considered in \cite{kim_07}. These calculations will be performed
whilst parameters in the model, such as reaction rates and cell volumes, are 
varied, to identify the impact of changing these parameters. To study these 
multi-cell models in a stochastic framework we will use some ideas from 
previous work, where the LNA was applied to multi-compartment biochemical models \cite{challenger_12}.

 Although we will be interested in a multi-cell model, we will begin with a 
one-cell model, which consists of a cell and an extra-cellular environment. 
The $cAMP$ within the cell is denoted by $cAMPi$, the $cAMP$ in the 
extra-cellular environment is denoted by $cAMPe$. The other components present 
in the model are adenylyl cylase (ACA), protein kinase (PKA), mitogen-activated
protein kinase (ERK2), the cAMP phosphodiesterase (RegA) and the ligand-bound 
cell receptor (CAR1). The coupling between the two compartments is caused by 
the external $cAMP$ binding to the cell receptor. To label the species we use: 
$(\textrm{ACA},\textrm{PKA},\textrm{ERK2},\textrm{RegA},\textrm{cAMPi},\textrm{cAMPe},\textrm{CAR1})=(x_1,x_2,x_3,x_4,x_5,x_6,x_7)$. The reaction scheme, 
together with the reaction kinetics and numerical values of the reaction rates are given in Table 1, 
which can be used to calculate the matrices $M$ and $B$ using Eq.~(\ref{MandB}).

\begin{widetext}
\begin{center}
\begin{table}[h]
\begin{center}
\begin{tabular}{|c|c|c|c|}
\hline
\multicolumn{4}{|c|} {Reaction Scheme} \\
\hline
Reaction & $f_{\mu}(\boldsymbol{x})$ & $\nu_{i\, \mu}$ & Rate Constant Value \\
\hline
$CAR1 \stackrel{k_1}{\longrightarrow} ACA + CAR1$ & $k_1x_7$ & $\nu_{1\,1}=+1$ & 
$k_1=1.96\textrm{min}^{-1}$ \\
\hline
$ACA + PKA \stackrel{k_2}{\longrightarrow} PKA$ & $k_2x_1x_2$ & $ \nu_{1\,2}=-1$ & 
$k_2=0.882\mu\textrm{M}^{-1}\textrm{min}^{-1}$ \\
\hline
$cAMPi \stackrel{k_3}{\longrightarrow} PKA + cAMPi$ & $k_3x_5$ & $\nu_{2\,3}=+1$ & 
$k_3=2.55\textrm{min}^{-1}$ \\
\hline
$PKA \stackrel{k_4}{\longrightarrow} \emptyset$ & $k_4x_2$ & $\nu_{2\,4}=-1$ &
$k_4=1.53\textrm{min}^{-1}$ \\
\hline
$CAR1 \stackrel{k_5}{\longrightarrow} ERK2 + CAR1$ & $k_5x_7$  & $\nu_{3\,5}=+1$ &
$k_5=0.588\textrm{min}^{-1}$ \\
\hline
$PKA + ERK2 \stackrel{k_6}{\longrightarrow} PKA$ &  $k_6x_2x_3$ & $\nu_{3\,6}=-1$ &
$k_6=0.816\mu\textrm{M}^{-1}\textrm{min}^{-1}$ \\
\hline
$\emptyset \stackrel{k_7}{\longrightarrow} RegA$ &  $k_7$ & $\nu_{4\,7}=+1$ &
$k_7=1.02\mu\textrm{M}\textrm{min}^{-1}$ \\
\hline
$ERK2 + RegA \stackrel{k_8}{\longrightarrow} ERK2$ &  $k_8x_3x_4$ & $\nu_{4\,8}=-1$ &
$k_8=1.274\mu\textrm{M}^{-1}\textrm{min}^{-1}$  \\
\hline
$ACA \stackrel{k_9}{\longrightarrow} cAMPi + ACA$ & $k_9x_1$ & $\nu_{5\,9}=+1$ &
$k_9=0.306\textrm{min}^{-1}$ \\
\hline
$RegA + cAMPi \stackrel{k_{10}}{\longrightarrow} RegA$ & $k_{10}x_4x_5$   & $\nu_{5\,\, 10}=-1$ &
$k_{10}=0.816\mu\textrm{M}^{-1}\textrm{min}^{-1}$  \\
\hline
$ACA \stackrel{k_{11}}{\longrightarrow} cAMPe + ACA$ & $k_{11}x_1$ & $\nu_{6\,\, 11}=+1$ &
$ k_{11}=0.686\textrm{min}^{-1}$  \\
\hline
$cAMPe \stackrel{k_{12}}{\longrightarrow} \emptyset$ &   $k_{12}x_6$   & $\nu_{6\,\, 12}=-1$ &
$k_{12}=4.998\textrm{min}^{-1}$  \\
\hline
$cAMPe \stackrel{k_{13}}{\longrightarrow} CAR1 + cAMPe$ & $k_{13}x_6 $ & $\nu_{7\,\,13}=+1$ &
$k_{13}=22.54\textrm{min}^{-1}$  \\
\hline
$CAR1 \stackrel{k_{14}}{\longrightarrow} \emptyset$ & $k_{14}x_7$  & $\nu_{7\,\,14}=-1$ &
$k_{14}=4.59\textrm{min}^{-1}$   \\
\hline
\end{tabular}
\caption{The reactions and their associated rate constants, given in micromoles and minutes. All the reactions have mass-action kinetics. Only non-zero entries of the stoichiometry matrix, $\nu_{i\mu}$, are shown.}
\label{table_1}
\end{center}
\end{table}
\end{center}
\end{widetext}

\begin{figure}
\begin{center}
\includegraphics[scale=0.9]{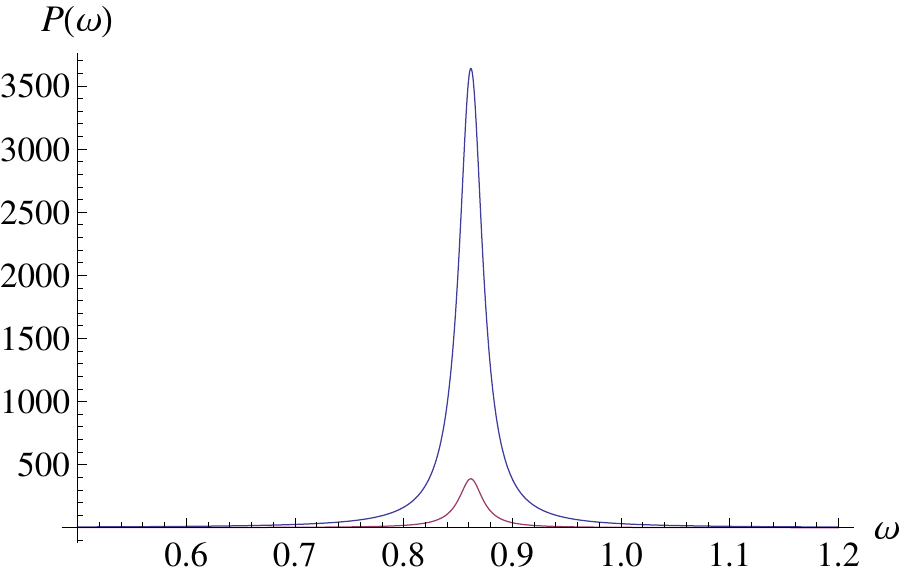}
\caption{Theoretical power spectra for the internal (larger peak) 
and external $cAMP$ in the one-cell model. The peak is around $\omega=0.83$. The reaction rates are
given in Table 1. In {\cite{kim_07}} the cell volume was chosen to be 
$3.672\times 10^{-14}\textrm{l}$. }
\label{fig:cAMP2}
\end{center}
\end{figure}

\begin{figure}
\begin{center}
\includegraphics[scale=0.9]{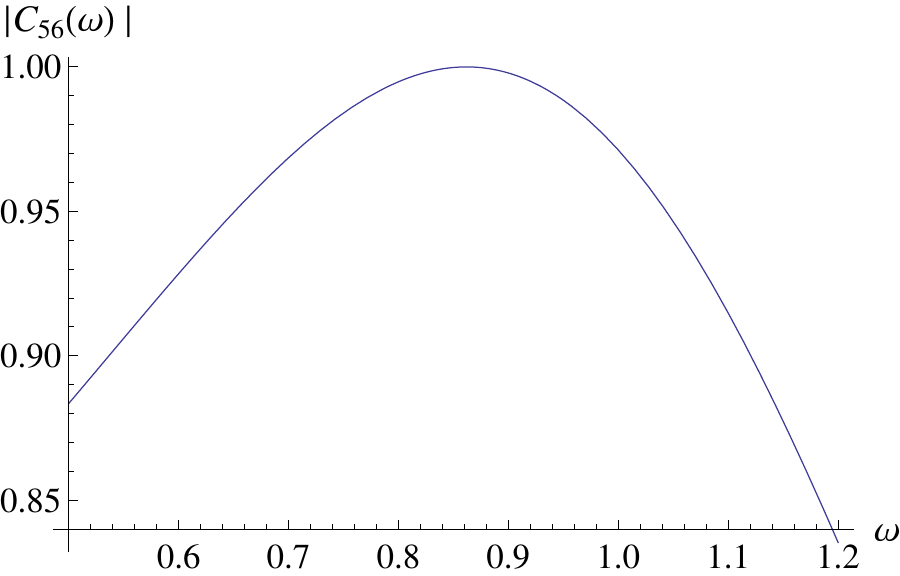}
\caption{Magnitude of the (theoretical) CCF for the internal and external 
cAMP fluctuations in the one-cell model. The reaction rates are
given in Table 1.}
\label{fig:cAMP3}
\end{center}
\end{figure}

\begin{figure}
\begin{center}
 \includegraphics[scale=0.9]{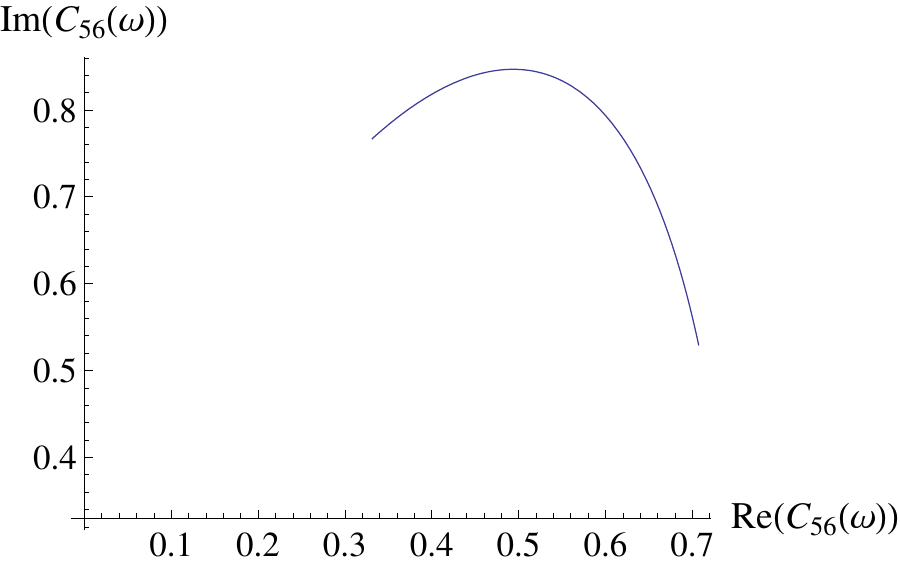}
\caption{Parametric plot of the (theoretical) CCF for the internal and 
external $cAMP$ fluctuations in the one-cell model. Only results for $0.5\leq \omega \leq 1.2$ 
are shown. The reaction rates are
given in Table 1.}
\label{fig:cAMP4}
\end{center}
\end{figure}

Some results from the one-cell model will be shown, before looking at the two 
cell model. The reaction parameters are chosen to be those which yielded the 
steady state in Figure 1B in {\cite{kim_07}}. As in Section 2, we use the linear
noise approximation to quantify the fluctuations around the steady state. We 
can then use Eq.~(\ref{psdm}) to calculate the PSDM and Eq.~(\ref{ccf}) to calculate the CCF. The specific form of the 
analytic results are not very illuminating and we content ourselves with 
showing the results graphically. Figure~\ref{fig:cAMP2} shows the power 
spectra for the internal and external $cAMP$. The values of the reaction 
parameters are given in the caption.  
Figure~\ref{fig:cAMP3} shows the magnitude of the CCF for these two species, 
showing a very strong correlation, especially in the frequency range within 
which the oscillations are significant. The CCF is displayed parametrically 
in Figure~\ref{fig:cAMP4}. It has both real and imaginary components, which indicates a phase lag. 

\begin{figure}
\begin{center}
 \includegraphics[scale=0.95]{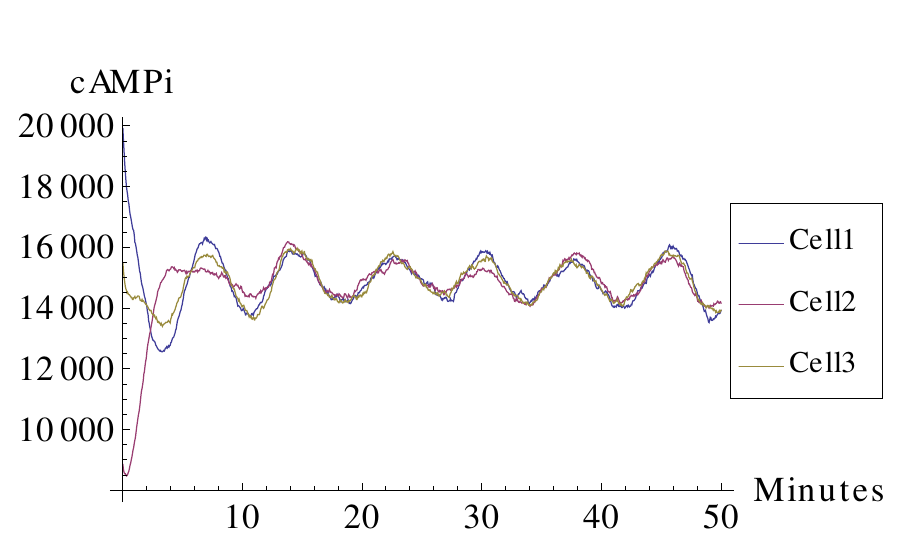}
\caption{Time series for $cAMPi$ (in particle numbers) in different cells for 
a three-cell model, generated using the Gillespie algorithm. The three species 
are given different initial conditions, but within a short amount of time they 
oscillate in phase with each other around their common fixed point.}
\label{fig:cAMP5}
\end{center}
\end{figure}

In {\cite{kim_07}}, the authors construct models of many cells, which are all 
coupled \textit{via} the external $cAMP$. A diagram of such a model, with 
three cells, is shown in Figure 5A of {\cite{kim_07}}. If the 
reaction parameters within each cell are identical, the oscillations in the 
three cells become synchronised with zero-phase lag. This effect can be easily 
seen by looking at the time series. This is shown for a three-cell model in 
Figure~\ref{fig:cAMP5}. To apply our analysis to a multi-cell model we proceed as we did for the one cell model: for an $n$ cell model there will be $6n+1$ variables.

\begin{figure}
\begin{center}
\includegraphics[scale=0.9]{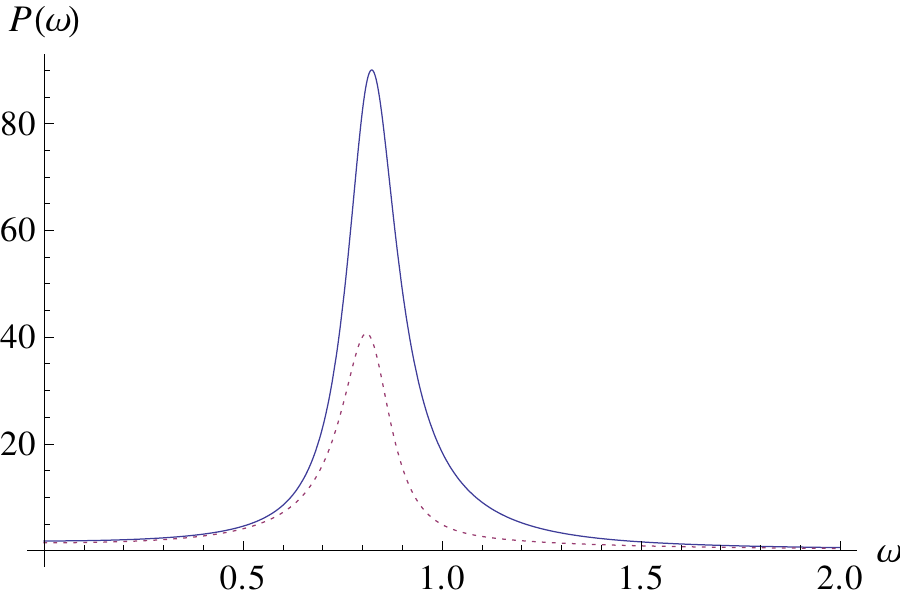}
\caption{Theoretical power spectra for $cAMPi$ in each cell of the two-cell 
model. The dashed line corresponds to the oscillations in the second cell, 
for which the reaction parameters are: $k_1=2.3\textrm{min}^{-1}$, $k_2=0.96\mu\textrm{M}^{-1}\textrm{min}^{-1}$, $k_3=2.1\textrm{min}^{-1}$, 
$k_4=1.9\textrm{min}^{-1}$, $k_5=0.4\textrm{min}^{-1}$, $k_6=0.72\mu\textrm{M}^{-1}\textrm{min}^{-1}$, $k_7=0.7\mu\textrm{M}\textrm{min}^{-1}$, 
$k_8=1.05\mu\textrm{M}^{-1}\textrm{min}^{-1}$, 
$k_9=0.26\textrm{min}^{-1}$, $k_{10}=0.89\mu\textrm{M}^{-1}\textrm{min}^{-1}$, $k_{11}=0.46\textrm{min}^{-1}$, 
$k_{13}=15\textrm{min}^{-1}$, $k_{14}=5.8\textrm{min}^{-1}$. The parameter 
$k_{12}$ was set to 9$\textrm{min}^{-1}$.}
\label{fig:cAMP6}
\end{center}
\end{figure}

\begin{figure}
\begin{center}
\includegraphics[scale=0.9]{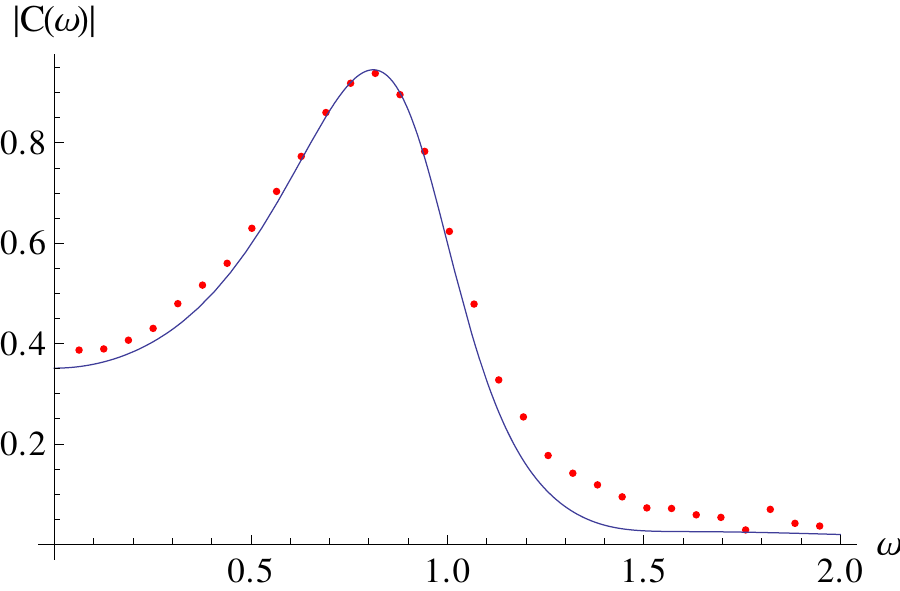}
\caption{The magnitude of the CCF for the $cAMPi$ in each cell of the two-cell 
model. The solid line is the theoretical result, the dots are from 6000 
numerical simulations}
\label{fig:cAMP7}
\end{center}
\end{figure}

\begin{figure}
\begin{center}
 \includegraphics[scale=0.9]{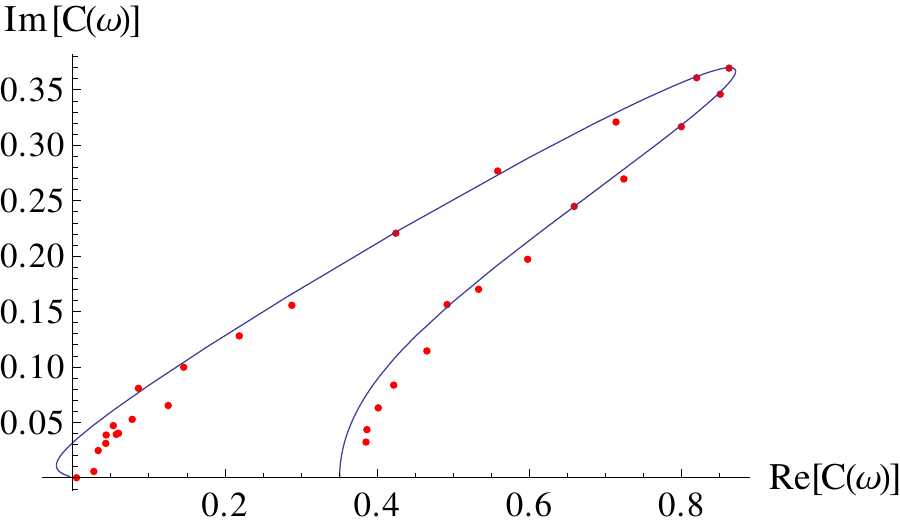}
\caption{Parametric plot of the CCF for the $cAMPi$ 
in each cell of the two-cell model. The solid line is the theoretical result, 
the dots are from 6000 numerical simulations. 
Results are shown for $0\leq \omega \leq 3$.}
\label{fig:cAMP8}
\end{center}
\end{figure}

\begin{figure}
\begin{center}
 \includegraphics[scale=0.9]{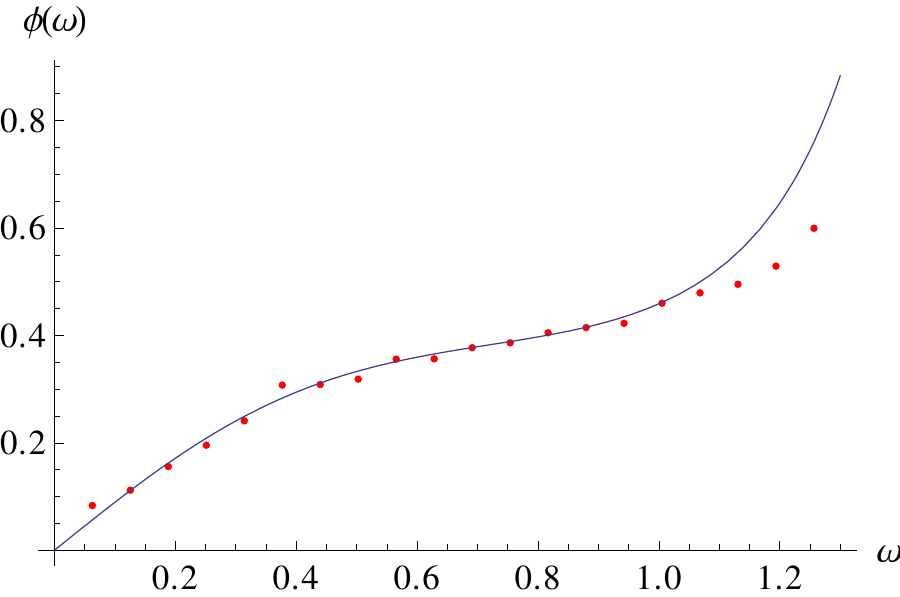}
\caption{Phase lag for oscillations of $cAMPi$ in each cell for the two-cell 
model. The solid line is the theoretical result, the dots are from 6000 
numerical simulations. Over the frequency range where oscillations are 
significant, the phase lag remains fairly constant.}
\label{fig:cAMP9}
\end{center}
\end{figure}

Synchronisation with a non-zero phase lag can be found when the reaction 
parameters differ from cell to cell. We illustrate this with a two cell model, 
where the reaction parameters in one cell are those used for the single cell 
model, and the reaction parameters in the second cell are obtained by making 
random perturbations to the original parameters. Below the relation between 
the internal $cAMP$ oscillations in each cell is examined, for one such 
parameter choice. Figure~\ref{fig:cAMP6} shows the power spectra for these 
two species. Figures~\ref{fig:cAMP7} \& \ref{fig:cAMP8} show the relevant CCF 
for these two species, whilst Figure~\ref{fig:cAMP9} displays the phase lag 
present, calculated using Eq.~(\ref{phase}). The theoretical results show good agreement with those from numerical simulation. Although the reaction parameters are different in each cell, there 
remains a strong shared signal in the oscillations. 

\begin{figure}
\centering
\subfigure{
 \includegraphics[width=3.35in]{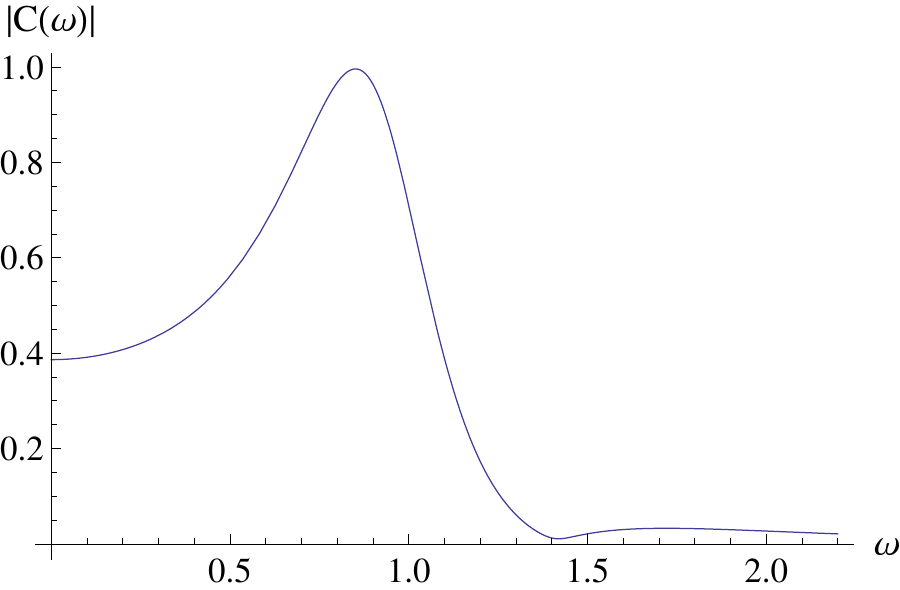}
}\\
\subfigure{
\includegraphics[width=3.35in]{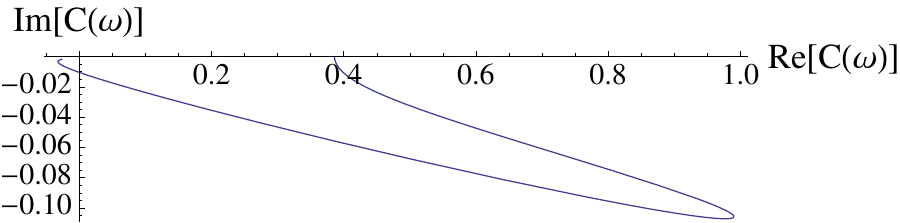}
}
\caption{Results for a two-cell model, where the cell volumes are not 
identical. Here, the cell volumes were chosen to be $V_I$ and $1.4V_I$, where 
$V_I$ is the cell volume chosen by Kim. \textit{et. al.}. The reaction 
parameters were the same in each cell, and were those used for the one-cell 
model. Top: The magnitude of the CCF for the $cAMP$ 
fluctuations in each cell. Bottom: A parametric plot, showing the real and 
complex parts of the same CCF for the frequency range 
$0\leq \omega \leq2$. A small phase lag has appeared, 
due to the different cell volumes.}
\label{fig:cAMP10}
\end{figure}

\begin{figure}
\centering
\subfigure{
 \includegraphics[width=3.35in]{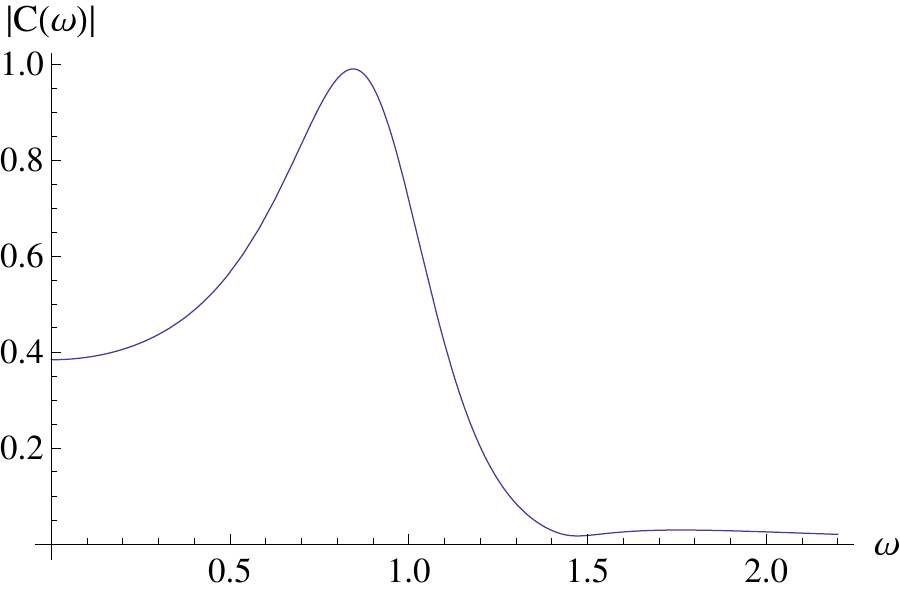}
}\\
\subfigure{
\includegraphics[width=3.35in]{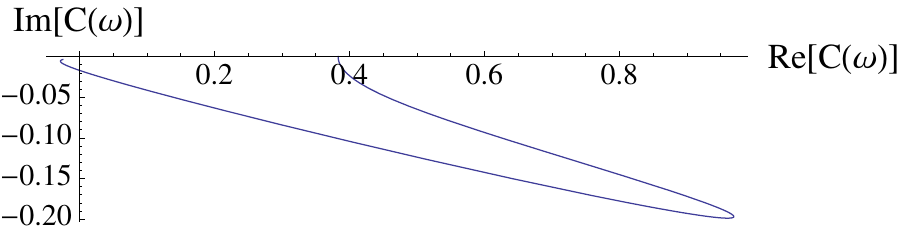}
}
\caption{Results for a two-cell model, where the cell volumes are not 
identical. Here, the cell volumes were chosen to be $V_I$ and $2V_I$, where 
$V_I$ is the cell volume chosen by Kim.~\textit{et.~al.}. The reaction 
parameters were the same in each cell, and were those used for the one-cell 
model. Top: The magnitude of the CCF for the $cAMP$ 
fluctuations in each cell. Bottom: A parametric plot, showing the real and 
complex parts of the same CCF for the frequency range 
$0\leq \omega \leq2$. A phase lag has appeared, due to the different 
cell volumes.}
\label{fig:cAMP11}
\end{figure}

\begin{figure}
\begin{center}
 \includegraphics[scale=0.9]{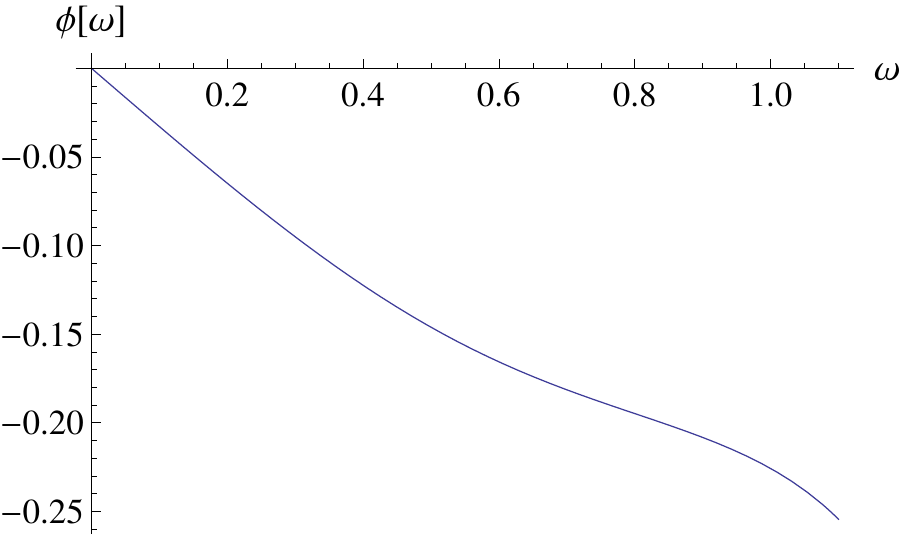}
\caption{The phase spectrum for the $cAMP$ fluctuations in each cell in a 
two-cell model for the case when the cell volumes are not identical. The 
cell volumes were $V_I$ and $2V_I$. The power spectra for these 
oscillations peak at $\omega=0.9$. The phase lag for oscillations of this 
frequency is 0.2 radians.
}
\label{fig:cAMP12}
\end{center}
\end{figure}

Throughout this section so far, we have assumed that all of the cells have 
the same volume. We will now look at the consequences of relaxing this 
restriction, which could be due to natural variation in the cell size. We find 
that a phase lag is introduced when the cells are of different sizes. Given 
that, for this particular system, it is desirable to have in phase 
synchronisation, it is important to see how large these phase lags are for 
various differences in cell size. To do this, we again studied a two-cell 
model, in which we fixed the volume of one cell, at the volume used previously
which we will call $V_I$, and the extracellular region (fixed at $5V_I$), and 
varied the volume of the other cell. The reaction parameters in each cell were 
identical and were chosen to be those used for the one cell model. The 
parameter governing the rate of degradation of $cAMPe$, $k_{12}$, was set to 
12$\textrm{min}^{-1}$. Figure~\ref{fig:cAMP10} shows details of the CCF for 
the case where the second cell has volume $1.4V_I$. The parametric plot shows 
that the imaginary component of the CCF is much smaller than the real 
component: this means that the phase lag is very small. We then increased the 
volume of the second cell to $2V_I$. Figure~\ref{fig:cAMP11} shows that the 
imaginary part of the CCF is larger than before, indicating a more significant 
phase lag. The phase spectrum for this system is shown in 
Figure~\ref{fig:cAMP12}. At the frequency for which the oscillations are most 
significant (the frequency for which the power is greatest), the phase lag is 
about 0.2 radians, roughly $11^\circ$. This shows that quite large differences 
in volume are required to introduce measurable phase lags: varying the cell 
size by a few percent does not have a significant effect. This is positive 
from the point of view of {in-phase} synchronisation being robust in this 
system. We also repeated this analysis whilst varying the volume of the 
extracellular region. However, this did not produce significantly different 
results from those presented here.

In this section we have studied a model of $cAMP$ oscillations in \textit{Dictyostelium}, due to Kim \textit{et. al}. Using the theoretical framework of the LNA, we have shown how analytical results can complement the numerical results which have already been obtained from this model \cite{kim_07}. 
The analysis performed in this section extends naturally to models containing many cells. However, in reality the \textit{Dictyostelium} cells can form aggregates containing thousands of cells. In this case, the distances between cells will be much larger, and the diffusion speed of the molecules would need to be taken into consideration. A spatial model would be required to capture this effect, although the mechanism by which the cells synchronise would be the same as the one outlined here.

\section{Discussion}
\label{sec:summary_sync}

In this work we have shown how stochastic oscillations in biochemical models 
can synchronise, and how analytical expressions for this effect may be found. 
Comparisons between numerical results and theoretical predictions for the CCF 
(and related quantities) were found to be good, especially in frequency ranges 
where the oscillations were significant. There have been very few models devised 
which study the synchronisation of stochastic oscillations across multiple 
cells, and those which have been constructed, such as {\cite{kim_07}}, are examined 
using numerical simulation. We hope that the analytical work described here 
can complement the numerical approach. In our study of the model of $cAMP$ 
signalling, we found that the stochastic oscillations across cells strongly 
influenced each other, and oscillations of species in different cells 
synchronised rapidly, as illustrated in Figure~\ref{fig:cAMP5}. In the case 
where the cells were of equal volume, and the reaction parameters across cells 
were the same, the oscillations synchronised in phase. Introducing changes to 
the reaction parameters in different cells, or making the cell volumes 
different to each other introduced a phase lag. These findings are similar to 
those reported in {\cite{rozhnova_12}}, in the context of epidemiological 
modelling. However, quite large changes had to be made before the lag was 
found to be significant. These results are positive for in-phase 
synchronisation being robust in this system. One slightly surprising result 
we found was that the size of the phase-lag was unaffected by varying the 
volume of the extracellular compartment. This is probably not realistic, and 
could be due to the assumption that all compartments are well mixed.  For a 
larger extracellular compartment, communication between cells would be 
expected to take longer, which could affect the phase lag. 

The analysis used to study the model of $cAMP$ oscillations was also applied 
to a model of oscillations in yeast glycolysis, due to Wolf \& Heinrich 
\cite{wolf_00}. Although not presented here, very different results were obtained for this model. 
The magnitude of the CCF for oscillators in different cells was extremely small. 
Using the reaction parameters chosen in the original article, we obtained 
CCFs with a magnitude of $\leq 0.01$ over the relevant frequency range. This 
indicates that the shared signal in the oscillators is extremely weak. The 
magnitude of the CCF remained small throughout an extensive parameter search of 
the model. This weak coupling between the cells could explain why, in the
deterministic framework, the oscillators take a longer-than-expected time to 
synchronise. The relation between the time to synchronise and the magnitude 
of the CCF could be an interesting avenue for further work. The magnitude of 
the fluctuations, here governed by the system-size, could also influence 
whether synchronisation is observed \textit{i.e.} if the system-size is small 
(and therefore the fluctuations are relatively large), is the synchronisation 
of two oscillators observed if the magnitude of the corresponding CCF has low 
or intermediate value? Such questions we leave for the future. Our aim here 
has been to formulate the analytic study of synchronisation of stochastic 
oscillators in a biochemical context, and to illustrate its use in a specific
model. We hope that this will encourage others use these techniques to 
investigate other biochemical systems.

\bigskip

\section*{Acknowledgements} JDC acknowledges support from the EPSRC and Program Prin2009 financed by the Italian MIUR.

\end{document}